\documentstyle{article}

\begin{document}
\vspace{2cm}
\begin{center}
\begin{large}
{\bf SCALINGS OF PITCHES IN MUSIC}
\end{large}

YU SHI

{\em Fudan-T. D. Lee Physics Laboratory and Department of Physics, Fudan
University, Shanghai 200433, PR China}
\end{center}
\vspace{2cm}

\begin{center}
{\bf Abstract}
\end{center}

We investigate correlations among pitches in several songs and pieces of
piano music by mapping them to one-dimensional walks.
Two kinds of correlations are studied, one is related to the  real
values of frequencies while they are treated only as different symbols
for another. Long-range power law behavior is found in both kinds. The
first is  more
meaningful. The structure of music, such as beat, measure and stanza, are
reflected in the change of scaling exponents. Some interesting features are
observed. Our results demonstrate the viewpoint that the fundamental
principle of music is the balance between repetition and  contrast.
\newpage

The aesthetic sense and perception remains a mystery in scientific
level. One may expect that its mechanism
should be dependent on the structure, especially a
kind of collective effect of information
possessed by the artistic or natural beauty and then the sensory processing
of this information. The quantitative nature of most elements of music make
analysis using methods of statistical physics feasible.
$1/f$ noise was discovered in an earlier measurement of loudness
fluctuations in
music$^{1}$. Motivated by what was done recently on DNA sequences and
human writings$^{2,3,4}$, we study
here  correlations among pitches according to musical scores.
25 songs
and several pieces of piano music are studied$^{5}$.

In equally tempered scale, the octave is divided into 12 semitones.
A semitone is the interval between two tones whose frequency ratio is the
twelfth root of 2. We consider two kinds of correlations, referred to as
``frequency-dependent'' (FD) one  and ``frequency-independent'' one
respectively. In the study of FD correlation, we map the real values of
tones to
positions within a one-dimensional walk, with a semitone  corresponding
to the unit of distance. The least of the duration of tones
in the piece of music studied
is  selected as the duration unit. The rests are neglected.
Hence the step of the walk, which corresponds to  the difference between
adjacent  pitches, is not a constant.
This walk reflects directly the change in the logarithm of frequency.
Consequently,  notes on the staff just make up an approximate landscape
of this so-called ``FD musical walk".

The correlation can be quantified by studying  the displacement
$\Delta X(t,T)$ of the walker after $T$ steps, which is the difference
between the positions $X(t+T)$ and $X(t)$,
\begin{equation}
\Delta X(t,T)\,=\,X(t+T)-X(t).
\end{equation}
$\Delta X(t,T)$ is just the sum of changes in each step, i.e.,
\begin{equation}
\Delta X(t,T)\,=\,\displaystyle\sum^{T-1}_{i=t}x(i),
\end{equation}
where
\begin{equation}
x(i)=X(i+1)-X(i)
\end{equation}
is the change in one step.
The mean square fluctuation which quantifies the correlation is
\begin{equation}
F_{FD}^{2}(T)\,=\,<[\Delta X(t,T)-<\Delta X(t,T)>_{t}]^{2}>_{t},
\end{equation}
where $<\,>_{t}$ means average over different $t$. It is well known$^{3}$
that for scaling signals,
\begin{equation}
F_{FD}^{2}(T)\,\sim\,T^{\alpha}.
\end{equation}
The corresponding power spectrum of $x$ and $\Delta X$ are respectively
$S_{x}(f)\,\sim\,f^{1-\alpha}$ and $S_{\Delta X}(f)\,\sim\,f^{-1-\alpha}$.
For an uncorrelated random walk, $\alpha$ = 1, $x$ exhibits white
noise while $\Delta X$ exhibits $1/f^{2}$ noise.
As $\alpha\,\rightarrow\,0$,
$x\,\rightarrow\,f$ noise and $\Delta X\,\rightarrow\,1/f$ noise.
As $\alpha\,\rightarrow\,2$,
$x\,\rightarrow\,1/f$ noise and $\Delta X\,\rightarrow\,1/f^{3}$ noise.

In the study of FI correlation, again the least duration of note is
selected as the unit. Pitches as well as rests   are treated just as
different symbols. They
are represented by different
numbers, which are then transformed to binary numbers. As in the studies on
DNA sequences and writings, ``0'' or ``1'' is mapped to up ($y(i)=+1$) or
down
($y(i)=-1$) in a one-dimensional walk, which can be termed
``FI musical walk''. The displacement is $\Delta
Y(t,T)\,=\,
\sum_{i=t}^{T-1}y(i)$. Similar to FD correlation, mean square fluctuations
$F_{FI}^{2}(T)$ of $\Delta Y$ are calculated. For scaling signals,
\begin{equation}
F_{FI}^{2}(T)\,\sim\,T^{\gamma}.
\end{equation}

As a typical example, Fig. 1(a) shows  the fluctuation in FD musical walk for
the national hymn of France, ``Marseillaise''. One
may observe the
periodicity due to strophic form with period $\approx\,400$. We may
only consider one, e.g., the first period (stanza). It is obvious that
there are three scaling regions, after the first with $\alpha_{1}\,
\approx\,1.029$
and the second with $\alpha_{2}\,\approx\,0.682$, there is a platform
region with $\alpha\,\approx\,0$. There is unavoidable
oscillation and deviation in the platform region, as well as break
down due to finite-size in large enough duration. This feature is found to be
universal
within 25 famous songs. Usually there are two to three regions with
nonzero $\alpha$'s before the final platform.

Quantitative results are summarized in Table 1.
 From the time and  duration unit
one may know how long a beat or a measure is. For instance, the unit of
``marseillaise'' is sixteenth, since the time is 4/4, the duration of a
beat is 4 while that of a measure is 16. Hence we see that the region I
and region II correspond to a beat and a measure respectively.  This
is valid for
most, though not all, of the songs studied. For
``Serenade''
by Schubert and ``God Save The Queen'', it can be thought that regions
I and
II merge to one. For
most of the songs, $\alpha_{1}$ is very close to $1$, indicating that
{\em usually the pitches within one beat are nearly random}.
That $\alpha_{2}$ and
$\alpha_{3}$ are smaller than $1$ indicates {\em
non-trivial correlations within
duration around a measure}, the power spectrum are between $1/f$ and
$1/f^{2}$. For longer duration, $\alpha\,\approx\,0$, indicating $1/f$ noise.
Therefore one may say that
 the longer the duration, the nearer to $1/f$ noise.

 We may make some other observations. The comparatively larger deviation
 from unity in $\alpha_{1}$ for three songs ``Auld lang Syne'', ``My old
 Kentucky home'' and ``Alamuhan'' show correlations even within one beat
 in a few songs.
 It is interesting to note that the platform begins immediately after
 one beat in the two ``Cradle song''s and in ``Santa Lucia'', which
 gives similar impression to cradle songs. There are also several other
 songs for which platform begin in duration shorter than one measure, while
 for some longer. We also notice on the plots for the two ``Serenade''s that
 the oscillation around the platform region particularly
 small. For several songs, such as ``Away in a Manger'', ``God Rest
 You Merry, Gentleman'',``Auld Lang Syne'', ``Old Black Joe'', ``Changing
 Partner'', ``Kangding Love Song'' and ``Alamuhan'',
 $\alpha$'s before the platform are comparatively close.

Fig. 1(b) shows fluctuation in FI musical walk for ``Marseillaise''.
After a fairly flat
non-power-law region, which is of little interest since it
corresponds to one or two duration unit, there is a region with non-zero
exponent $\gamma$,
then crossover to a platform region with $\gamma\,\approx\,0$. Periodicity
due to strophic form is also indicated. The structure indicationg
more than one scaling region before the platform in the FD
correlation has not been observed in FI correlation. This feature
is universal for all the songs. The values of $\gamma$'s are listed in the
last column of Table 1.  All $\gamma$'s are larger than 1. It
should be noticed that FD fluctuation is for the move of pitch,
while FI fluctuation is for the {\em summation}
of binary representation of pitches. If $\gamma$ were unity, the pitches
would
be white noise. $1<\,\gamma\,<\,2$ indicates that the pitches are between
white
noise and $1/f$ noise. $\gamma\,\approx\,0$ indicates that the pitches are
nearly
$f$ noise while their summations are $1/f$ noise. Therefore, long range
correlations also exhibit in FI correlation.

The same schemes are also used to treat several pieces of piano music, both
melody and accompaniment are analyzed.
``Courante''s and ``Sarabande''s in 4th ``English Suite'', 5th and 6th
``French Suite''s by J. S. Bach and ``Variation (G major)'' by Beethoven
are studied. The general feature is similar to that for the songs, apart
from that there is oscillation modulation before the final platform region in
the FD
fluctuation
of ``Courante''s in ``French Suite''s and in Var. VI in ``Variation'',
as shown
in Fig. 2.
The quantitative results are summarized in Table 2. We may
observe comparative similarities
between the two in the binary form, between melody and accompaniment,
between ``Courante''s or ``Sarabande''s
in different French Suites, as well as between different parts of
one Suite.  For FD correlation, it
can be observed that the platform begins earlier than songs. For every
piece in ``English Suite'',
it begins just after one beat. For most of
the others, it begins in duration not longer than one measure. Apart from
three, ``$\gamma$''s  are also larger than unity.

Given the scalings, we may renormalize the music by selecting out pitches
with a certain distance of duration while neglecting others. As an
example, the FD correlation of renormalized ``Marseillaise''
is shown in Fig. 1(c). It is clear that the scaling behavior remains
unchanged
while the turning  or crossover durations at which scaling changes are
reduced by the scaling factor. It is
interesting to have renormalized music performed and compared with the
original one.

The scaling behavior uncovered here demonstrates quantitatively
 the viewpoint
that the fundamental principle of music is the balance between repetition
and contrast$^{6}$.  Since the pitch corresponds to logarithm of frequency,
our results suggest that it is the logarithm of frequency that matters
in processing musical information in human brain. This conjecture is supported
by the finding$^{7}$  that the smallest perceptible frequency change increases
as the frequency change. This can be understood as that the
perceptible change in logarithm of frequency is constant.

In closing this article, we would like to point out
that many of the other elements
of music, such as chord, rhythm, timbre and dynamics etc., have not been taken
in account here. Further studies are anticipated concerning both music itself
and information processing on it.

\noindent
{\bf ACKNOWLEDGEMENTS}

The author is grateful to  Prof. Rui-bao Tao for much help. Ping Xia,
Beng-dao Shi, De-ping Wang and Chao Tang are thanked
for valuable discussions.

\noindent
{\bf REFERENCES}

\begin{enumerate}
\item R. F. Voss \& J. Clarke, {\em Nature} {\bf 258}, 317 (1975).

\item C. K. Peng etc. {\em Nature} {\bf 356} 168 (1992).

\item R. F. Voss, {\em Phys. Rev. Lett.} {\bf 68}, 3805 (1992).

\item A. Schenkel, J. Zhang \& Y. C. Zhang,  {\em Fractals} {\bf 1}, 47 (1993)

\item The references for musical scores we used are: {\em Going Places in
Harmony}, ed. R. Bobley, (Bobley, New York, 1976); {\em
111 Golden British and American Songs}, eds. M. Du and C. Yan, (Shanghai
Education, Shanghai, 1991); {\em 101 American Pop Songs}, eds. J. Yan and
D. Li, (Shanghai Translation, Shanghai, 1987); {\em Chinese Folk Songs},
ed. M. Ding, (New World, Beijing, 1984); {\em 100 Pieces of Piano Music},
ed. W. Ge et al., (Shanghai Music, Shanghai, 1994); J. S. Bach,
{\em English Suites}, (Shanghai Music, Shanghai, 1994); J. S. Bach,
{\em French Suites}, (Shanghai Music, Shanghai, 1994).

\item D. T. Politoske, {\em Music} (Prentice-Hall Inc., New Jersey, 1988).

\item C. Taylor,  {\em Exploring Music}
(IOP publishing, Bristol and Philadelphia, 1992).
\end{enumerate}
\newpage
Figure captions:

\noindent
Fig. 1 Double logarithmic plots of mean square fluctuations as a function
of the duration $T$
 for the
song ``Marseillaise''. (a) Frequency-dependent correlation.
(b) Frequency-independent correlation.
(c) Frequency-dependent correlation for this song  renormalized by
factor 2.
\vspace{1.5cm}

\noindent
Fig. 2 Mean square fluctuation of frequency-dependent musical
walk for the melody of theme B of the ``Courante'' in ``5th French
Suite''.  This kind of oscillation modulation structure before the platform
is found in
several piano pieces other than the most with simple power law scalings.

\newpage
\begin{small}
{\bf Table 1 Scaling regions and exponents of mean square fluctuations
of musical walks for 25 songs.
For each song, author or origin, key and time are indicated in the following
the  title. Keys are all major, unit is
that of duration. Reg. I, II and III gives
the different scaling regions before the final platform region for the
frequency-dependent correlation, $\alpha_{1}$, $\alpha_{2}$ and
$\alpha_{3}$ are the corresponding scaling exponents. $\gamma$
is the scaling exponent for the frequency-independent correlation.}

\vspace{0.8cm}
\begin{tabular}{lcccccccc}
\hline
song&unit&reg. I&$\alpha_{1}$&reg. II&$\alpha_{2}$
&reg. III&$\alpha_{3}$&$\gamma$\\     \hline \\
National Hymn of PRC (E. Nie, G, 2/4)&1/16&1$\sim$ 4&0.936&5$\sim$ 8&0.623&&
&1.199\\
God Save The Queen (UK, A, 3/4)&1/8&1$\sim$6&1.103&&&7$\sim$17&0.987&1.656\\
Marseillaise (R. de L'isle, G, 4/4)&1/16&1$\sim$4&1.029&4$\sim$16&0.682
&&&1.567\\
Star-spangled Banner (J.S. Smith, $^{b}$B, 3/4)&1/16&1$\sim$4&1.019
&4$\sim$8&0.835
&8$\sim$18&0.638&1.584\\
Silent Night (F. Gruber, C, 3/4)&1/8&1$\sim$3&1.083&4$\sim$10&0.665&&&1.724\\
Internationale (French, $^{b}$B, 4/4)&1/16&1$\sim$4&0.973&4$\sim$8&0.647
&9$\sim$12&0.533&1.548\\
O,Come,All Ye Faithful (J.F.Wade, A, 4/4)&1/8&1$\sim$2&1.013&2$\sim$6&0.620
&7$\sim$12&0.439&1.517\\
Joy to The World (G.F.Handel, D, 2/4)&1/16&1$\sim$4&1.044&4$\sim$8&0.766
&8$\sim$19&
0.424&1.561\\
Away in a Manger (J.R.Murray, F, 3/4)&1/8&1$\sim$2&1.107&2$\sim$4&0.929
&5$\sim$12&
0.925&1.667\\
Deck The Hall (Old Welsh Air, F, 4/4)&1/8&1$\sim$2&1.088&3$\sim$7&0.705&&
&1.144\\
God Rest You Merry, Gentleman (English, G, 4/4)&1/8&1$\sim$2&0.999&
3$\sim$8&0.945&&&1.156\\
Cherry Ripe (English folk song, $^{b}$E, 4/4)&1/16&1$\sim$3&0.969&4$\sim$16
&0.588&&&1.200\\
Auld Lang Syne (Scottish, F, 2/4)&1/16&1$\sim$3&0.863&4$\sim$8&0.834
&&&1.239\\
When You and I Were Young, Maggie (American, F, 4/4)&1/16&1$\sim$4&1.026
&5$\sim$8&
0.703&&&1.564\\
Santa Lucia (Neapolitan, C, 3/8)&1/32&1$\sim$2&0.983&2$\sim$4&0.640&&&1.566\\
Old Black Joe (S. Foster, D, 4/4)&1/16&1$\sim$8&0.937&9$\sim$12&0.824&&
&1.608\\
My Old Kentucky Home (S. Foster, G, 4/4)&1/16&1$\sim$4&0.874&4$\sim$8&0.462&&
&1.444\\
Cradle Song(J. Brahms, F, 3/4)&1/8&1$\sim$2&0.993&&&&&1.075\\
Cradle Song (F. Schubert, $^{b}$A, 4/4)&1/16&1$\sim$3&1.031&&&&&1.368\\
Serenade (F. Schubert, F, 3/4)&1/48&1$\sim$18&0.981&&&19$\sim$25&0.598
&1.703\\
Serenade (C. Gounod, F, 6/8)&1/48&1$\sim$6&0.992&7$\sim$36&0.429&&&1.511\\
Changing Partner
(American, A, 3/4)&1/8&1$\sim$6&0.974&7$\sim$11&0.959&&&1.268\\
Feelings (M.
Albert, G, 4/4)&1/48&1$\sim$12&0.979&13$\sim$17&0.686&18$\sim$40&0.438
&1.811\\
Kangding Love Song (folk song in Xikang of China, F, 4/4)&1/8&1$\sim$4&0.967
&5$\sim$7&0.883&&&1.519\\
Alamuhan (folk song in Uygur of China, F,
4/4)&1/16&1$\sim$4&0.757&5$\sim$8&0.689
&&&1.314\\ \hline
\end{tabular}

\newpage
\noindent
{\bf Table 2
Scaling regions and exponents of mean square fluctuations
of musical walks for several pieces of piano music. All keys are
major.
For ``Courante'' and ``Sarabande''  of a same Suite, keys are the
same hence indicated after the
the author Bach,
times are different hence indicated separately for
``Courant'' and ``Sarabande''.
Each piece in Suites is of binary form with two parts
A and B. There is oscillation modulation
before the platform region  in the
frequency-dependent
correlations of ``Courante''s in ``French Suites'', hence no quantitative
analysis is made for them. All parts of ``Variation'' have the same key and
time.}

\vspace{0.8cm}
\begin{tabular}{lcccccccc}
\hline
piano music&unit&reg. I&$\alpha_{1}$&reg. II&$\alpha_{2}$&reg. III
&$\alpha_{3}$&$\gamma$\\
\hline\\
 4th English Suite (J. S. Bach, F)\\
\hspace{10pt} Courante(3/2)&\\
\hspace{20pt} melody A&1/32&1$\sim$4&0.964&5$\sim$8&0.883
&9$\sim$16&0.595&0.959\\
\hspace{20pt}  melody B&1/32&1$\sim$4&1.005&5$\sim$8&0.805
&9$\sim$16&0.703&1.507\\
\hspace{20pt}  accompaniment A&1/32&1$\sim$4&1.001&5$\sim$8&
0.893&9$\sim$16&0.269&1.141\\
\hspace{20pt}  accompaniment B&1/32&1$\sim$4&1.002&5$\sim$8&0.964
&9$\sim$16&0.425&1.201\\
\hspace{10pt}   Sarabande(3/4)&\\
\hspace{20pt} melody A&1/32&1$\sim$4&0.911&5$\sim$8&0.826&9$\sim$16
&0.925&1.083\\
\hspace{20pt} melody B&1/32&1$\sim$4&0.958&5$\sim$8&0.755&9$\sim$16
&0.643&1.572\\
\hspace{20pt} accompaniment A&1/32&1$\sim$16&1.012&&&&&1.273\\
\hspace{20pt} accompaniment B&1/32&1$\sim$8&1.016&9$\sim$16&0.715&&&1.547\\
 \hspace{1cm}\\
 5th French Suit (J. S. Bach,G)&
\\
\hspace{10pt}   Courante\\
\hspace{20pt} melody A&&&&&&&&1.487\\
\hspace{20pt} melody B&&&&&&&&1.144\\
\hspace{20pt} accompaniment A&&&&&&&&1.582\\
\hspace{20pt} accompaniment B&&&&&&&&1.600\\
\hspace{10pt}  Sarabande(3/4)& \\
\hspace{20pt} melody A&1/32&1$\sim$4&0.938&5$\sim$8&0.839&9$\sim$12&0.632
&1.589\\
\hspace{20pt} melody B&1/32&1$\sim$5&0.914&6$\sim$22&0.675&&&1.581\\
\hspace{20pt} accompaniment A&1/16&1$\sim$4&0.972&4$\sim$6&0.708&6$\sim$12
&0.411&1.206\\
\hspace{20pt} accompaniment B&1/16&1$\sim$4&0.950&4$\sim$6&0.526&6$\sim$8
&0.352&1.099\\
\hline
\end{tabular}
\newpage
\begin{center}
{\bf Table 2} (Continued)
\end{center}
\vspace{0.8cm}

\begin{tabular}{lcccccccc}
\hline
\\
piano music&unit&reg. I&$\alpha_{1}$&reg. II&$\alpha_{2}$&reg. III
&$\alpha_{3}$&$\gamma$\\
 \hline\\
 6th French Suite (J. S. Bach, E)&\\
\hspace{10pt}   Courante()\\
\hspace{20pt} melody A&&&&&&&&1.292\\
\hspace{20pt} melody B&&&&&&&&1.588\\
\hspace{20pt} accompaniment A&&&&&&&&1.676\\
\hspace{20pt} accompaniment B&&&&&&&&1.332\\
\hspace{10pt}  Sarabande(3/4)&\\
\hspace{20pt}  melody A&1/16&1$\sim$8&0.769&9$\sim$12&0.471&&&1.668 \\
\hspace{20pt}  melody B&1/16&1$\sim$7&0.848&8$\sim$12&0.676&&&1.620 \\
\hspace{20pt}  accompaniment A&1/32&1$\sim$8&1.024&&&&&1.693 \\
\hspace{20pt}  accompaniment B&1/32&1$\sim$8&1.043&9$\sim$32&0.908&&&1.687 \\
\hspace{1cm}\\

Variation (L. v. Beethoven, G, 3/4)\\
\hspace{10pt}  melody\\
\hspace{20pt} Theme&1/32&1$\sim$4&1.005&4$\sim$8&0.666&&&0.854\\
\hspace{20pt} Var. I&1/32&1$\sim$2&1.003&3$\sim$6&0.918&&&1.338\\
\hspace{20pt} Var. II&1/32&1$\sim$4&1.001&4$\sim$8&0.308&&&1.219\\
\hspace{20pt} Var. III&1/32&1$\sim$4&0.946&4$\sim$6&0.615&&&1.263\\
\hspace{20pt} Var. IV&1/32&1$\sim$4&1.004&4$\sim$8&0.563&&&1.112\\
\hspace{20pt} Var. V&1/96&1$\sim$16&0.997&17$\sim$40&0.800&&&1.503\\
\hspace{10pt}   accompaniment\\
\hspace{20pt}  Theme&1/32&1$\sim$4&1.004&4$\sim$8&0.273&&&0.822\\
\hspace{20pt}  Var. I&1/32&1$\sim$4&1.005&4$\sim$8&0.155&&&1.118\\
\hspace{20pt}  Var. II&1/32&1$\sim$2&1.003&3$\sim$19&0.938&&&1.049\\
\hspace{20pt}  Var. III&1/32&1$\sim$4&1.014&4$\sim$6&0.722&&&1.295\\
\hspace{20pt}  Var. IV&1/32&1$\sim$4&1.015&4$\sim$8&0.737&&&1.355\\
\hspace{20pt}  Var. V&1/32&1$\sim$8&1.021&9$\sim$20&0.931&&&1.143\\
\hline
\end{tabular}
\end{small}
\end{document}